\documentclass[iop]{emulateapj}
\slugcomment{To appear in ApJ Letters}

\usepackage[svgnames,hyperref]{xcolor}
\usepackage{hyperref}
\hypersetup{colorlinks=True,linkcolor=Black,citecolor=Black,
urlcolor=Black,pagebackref}

\usepackage{nicefrac}

\usepackage{ amssymb }

\newcommand{\nustar}{\textit{NuSTAR}\ }
\newcommand{\swift}{\textit{Swift}\ }
\newcommand{\chandra}{\textit{Chandra}\ }
\newcommand{\nustarn}{\textit{NuSTAR}}
\newcommand{\swiftn}{\textit{Swift}}
\newcommand{\chandran}{\textit{Chandra}}
\newcommand{\grb}{GRB\,130925A\ }
\newcommand{\grbn}{GRB\,130925A}

\newcommand{\spm}[2]{\ensuremath{^{+#1}_{-#2}}}
\newcommand{\ppm}{\ensuremath{\pm}}

\shorttitle{X-ray Spectral Components in the Afterglow of GRB\,130925A}
\shortauthors{Bellm et al.}

\begin{document}


\title{X-ray Spectral Components Observed
in the Afterglow of GRB\,130925A}


\author{Eric C. Bellm\altaffilmark{1},
Nicolas M. Barri\`{e}re\altaffilmark{2},
Varun Bhalerao\altaffilmark{3},
Steven E. Boggs\altaffilmark{2}, 
S. Bradley Cenko\altaffilmark{4},
Finn E. Christensen\altaffilmark{5},  
William W. Craig\altaffilmark{2,6},  
Karl Forster\altaffilmark{1},
Chris L. Fryer\altaffilmark{7},
Charles J. Hailey\altaffilmark{8}, 
Fiona A. Harrison\altaffilmark{1}, 
Assaf Horesh\altaffilmark{9},
Chryssa Kouveliotou\altaffilmark{10},
Kristin K. Madsen\altaffilmark{1},
Jon M. Miller\altaffilmark{11},
Eran O. Ofek\altaffilmark{9},   
Daniel A. Perley\altaffilmark{1},
Vikram R. Rana\altaffilmark{1},
Stephen P. Reynolds\altaffilmark{12},
Daniel Stern\altaffilmark{13}, 
John A. Tomsick\altaffilmark{2},
and William W. Zhang\altaffilmark{4}
}

\altaffiltext{1}{Cahill Center for Astronomy and Astrophysics, California Institute of Technology, Pasadena, CA 91125; ebellm@caltech.edu.  }
\altaffiltext{2}{Space Sciences Laboratory, University of California, Berkeley, CA 94720}
\altaffiltext{3}{Inter-University Center for Astronomy and
Astrophysics, Post Bag 4, Ganeshkhind, Pune 411007, India}
\altaffiltext{4}{NASA Goddard Space Flight Center, Greenbelt, MD 20771}
\altaffiltext{5}{DTU Space - National Space Institute, Technical
University of Denmark, Elektrovej 327, 2800 Lyngby, Denmark}
\altaffiltext{6}{Lawrence Livermore National Laboratory, Livermore, CA 94550}
\altaffiltext{7}{CCS-2, Los Alamos National Laboratory, Los Alamos, NM 87545}
\altaffiltext{8}{Columbia Astrophysics Laboratory, Columbia University, New York, NY 10027}
\altaffiltext{9}{Benoziyo Center for Astrophysics, Weizmann Institute of Science, 76100 Rehovot, Israel.}
\altaffiltext{10}{Astrophysics Office/ZP12, NASA Marshall Space Flight Center, Huntsville, AL 35812, USA}
\altaffiltext{11}{Department of Astronomy, The University of Michigan, 500 Church Street, Ann Arbor, MI, 48109}
\altaffiltext{12}{Physics Dept., NC State U., Raleigh, NC 27695}
\altaffiltext{13}{Jet Propulsion Laboratory, California Institute of Technology, Pasadena, CA 91109}







\begin{abstract}

We have identified spectral features in the late-time X-ray afterglow of
the unusually long, slow-decaying \grb using \nustarn, \swiftn-XRT, and
\chandran.   A spectral component in addition to an absorbed power-law is
required at $>4\sigma$ significance, and its spectral shape varies between
two observation epochs at $2\times10^5$ and $10^6$ seconds after the burst.
Several models can fit this additional component, each with very different
physical implications.  A broad, resolved Gaussian absorption feature of
several keV width improves the fit, but it is poorly constrained in the
second epoch.  An additive black body or second power-law component provide
better fits.  Both are challenging to interpret: the blackbody radius is
near the scale of a compact remnant ($10^8$\,cm), while the second powerlaw
component requires an unobserved high-energy cutoff in order to be
consistent with the non-detection by \textit{Fermi}-LAT.

\end{abstract}




\section{Introduction}

Recent work has identified several ``ultra-long'' gamma-ray bursts (GRBs)
with properties distinct from normal long GRBs \citep[and references
therein]{tmp_Levan:13:ULGRBs}.  These events have initial bursting phases
lasting thousands of seconds in gamma-rays 
and show long-lived, highly variable X-ray afterglows.  It is
currently unclear whether these bursts are simply extreme examples of the
long GRB class, as suggested by \citet{tmp_Zhang:13:ULGRBs}; 
if they are related to the even
longer candidate relativistic Tidal Disruption Events (TDEs) \swift J1644+57
\citep{Bloom:11:SwiftJ1644TDF,Levan:11:SwiftJ1644TDF,
Burrows:11:SwiftJ1644TDF} and \swift J2058+05
\citep{Cenko:12:TDESwiftJ2058}; or if they represent a new subclass of
transient, perhaps with large-radius progenitors
\citep{Woosley:12:LongCollapsars,Gendre:13:ULGRB111209A,Nakauchi:13:BSGULGRB}.

The bright, nearby \grb is similar to previously reported ultra-long 
GRBs and, with the launch of \nustarn, provides an opportunity to  
observe the X-ray spectrum at high sensitivity over a broad energy band.
Here we report time-varying spectral features 
in the late-time X-ray afterglow of \grb that were initially discovered by
\nustar and confirmed in a second epoch by \nustar and \chandran.
Our detections
are at higher energies and significantly 
later times than previously reported afterglow features.

Before the era of routine afterglow observations with \swiftn-XRT, several
authors claimed detection of lines in GRB X-ray afterglows on top of
otherwise smooth power-law spectra
\citep[e.g.,][]{Piro:2000:AGLine,Amati:00:AGAbsLine,Reeves:02:AGLine}.
Most reports were of emission lines at relatively low signal-to-noise
ratio, and there was substantial controversy over the methods used to
assess line significance
\citep{Protassov:02:LRTest,
Sako:05:AGLineReanalysis}.
Since the advent of \swiftn, no firm afterglow line detections have been
reported despite its greater sensitivity and systematic followup, calling
previous reports into question \citep[for a review,
see][]{Hurkett:08:XRTLineSearches}.  

However, statistically significant blackbody components have been reported
in the early-time ($t \lesssim 10^3$\,s) 
afterglow spectra of several bursts observed by
\swiftn-XRT \citep[][and references therein]{Starling:12:AGBBody}.  
The inferred rest-frame temperatures are typically a few tenths of a keV, the
inferred radii are $\sim10^{12}$\,cm, and the blackbody component provides
10--50\% of the 0.3--10\,keV flux.
The first detections were in low-luminosity, SN-associated GRBs,
leading to suggestions that the emission was due to shock breakout from the
SN \citep[e.g.,][]{Campana:06:ThermalAG}.  
Systematic searches have found thermal components
in early afterglows of classical GRBs as well
\citep{Sparre:12:AGBBSearch,Friis:13:AGBBSearch}, 
giving credence to alternative
interpretations including 
late-time emission from a prompt photosphere \citep{Friis:13:AGBBSearch} or
emission from a cocoon around the jet
\citep{Suzuki:13:ThermalCocoon,Nakauchi:13:BSGULGRB}.

Of particular relevance are reports of additional components in the
afterglows of other ultra-long GRBs. The ``Christmas Day Burst''
GRB\,101225A showed evidence of two separate blackbody components, a 1\,keV
X-ray black body with radius $2\times10^{11}$\,cm observed 
6\,ksec after the burst 
and a UVOIR black
body with radius 2--7$\times10^{14}$\,cm which cooled over 18\,days
\citep{Thone:11:ChristmasDayBurst}.  In GRB\,111209A,
\citet{Stratta:13:ULGRB111209A} reported the \textit{XMM} detection of a
second, hard power-law component ($\Gamma \sim 0$) during the steep decay
phase $\sim$70\,ksec after the burst.

\section{Observations}

\grb produced several emission episodes triggering \swiftn-BAT,
\textit{Fermi}-GBM, and \textit{MAXI}.  \swiftn-BAT triggered on \grb at
$T_0 = $ 2013-09-25 04:11:24 UT \citep{Lien:2013:GCN15246}.  
\textit{Fermi} GBM triggered on a precursor episode about 15 minutes before
the \swift trigger \citep{Fitzpatrick:2013:GCN15255}, and \textit{MAXI}
triggered on an emission episode nearly 4\,ksec after the initial \swift
trigger \citep{Suzuki:2013:GCN15248}.  The final BAT detection of the
emission occurred during a flare observed by XRT, at $T_0+7.1$\,ksec
\citep{Markwardt:2013:GCN15257}.  Despite an automated repointing,
\textit{Fermi}-LAT did not detect any emission
\citep{Kocevski:2013:GCN15268}.  Both the \textit{INTEGRAL}-SPI
Anti-Coincidence Shield and
\textit{Konus-WIND} detected gamma-rays from the burst over a total
interval of nearly 5\,ksec
\citep{Savchenko:2013:GCN15259,Golenetskii:2013:GCN15260}.

\begin{figure}
\includegraphics[width=\columnwidth]{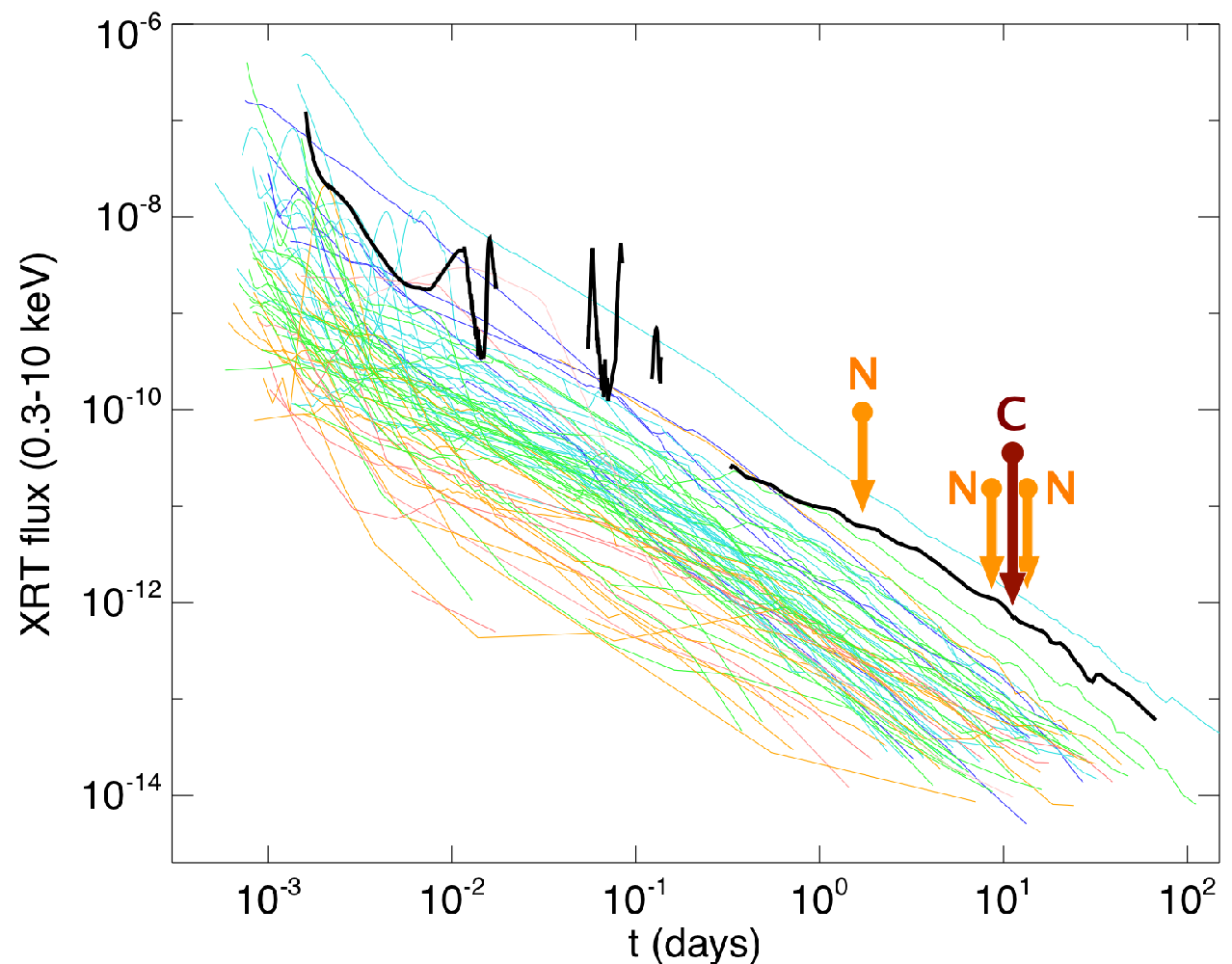}
\caption{\textit{Swift}-XRT lightcurve 
for \grb (black) plotted over the XRT lightcurves of other afterglows.
The \nustar (N) and
\chandra (C) observation times are marked.
\label{fig:lc}}
\end{figure}

\swiftn-XRT observed large, repeated flares from the burst
(\citealp{Evans:2013:GCN15254}; Figure \ref{fig:lc}).  
The extraordinary length of the bursting
phase led \citet{Burrows:2013:GCN15253} to suggest similarity to the proposed
jetted TDE \swift J1644+57, although
\citet{Golenetskii:2013:GCN15260}
argued that some previous ultra-long events thought to be GRBs
had been observed with similar total duration.

Starting around $2\times10^4$\,sec after the \swift trigger, the X-ray
afterglow entered a steady decay phase without new
flares (Figure \ref{fig:lc}).  The observed decline is similar to other GRB
afterglows and differs markedly from the weeks of flaring observed for \swift
J1644+57.  

In contrast, the source was faint at optical--NIR wavelengths.
Rapid followup observations found a NIR-bright ($K=18$, $r' > 22$\,mag AB)
source near the X-ray position \citep{Sudilovsky:2013:GCN15247}.  
Spectroscopy of the host galaxy
provided a redshift of $z=0.347$ \citep{Vreeswijk:2013:GCN15249,
Sudilovsky:2013:GCN15250}.  
Late-time \textit{HST} imaging showed that the event took place in the plane of a
disrupted host galaxy but offset 0.12\,arcsec (600\,pc in projection) 
from the galaxy nucleus
\citep{Tanvir:13:GCN15489}.  This offset disfavors a TDE origin for this
event, although the authors noted that a galaxy merger
could produce a supermassive black hole offset from the light centroid.

\nustar \citep{Harrison:2013:NuSTAR} provides unprecedented X-ray sensitivity
above 10\,keV thanks to the combination of its multilayer-coated focusing 
optics and CdZnTe detectors.
\nustar observed \grb during the decay phase
beginning 1.8 days after the \swift trigger (Figure \ref{fig:lc}).
The total on-source observation time in the first epoch
was 39.2\,ksec.
Our initial analysis showed that an absorption feature was needed to fit
the \nustar data \citep{Bellm:13:GCN15286}.
We triggered two additional \nustar observations of 88.2 and
90.7\,ksec integration time; these occurred at 8.8 and 11.3 days after
the \swift trigger.  We also obtained a 44.3\,ksec Director's Discretionary
Time observation with \chandra ACIS-S 
beginning 11.0 days after the \swift trigger.  

\section{Data Reduction}

We processed the \nustar data with HEASOFT 6.14 and 
the \nustar Data Analysis Software (NuSTARDAS) v.1.2.0 using CALDB
version 20130509.  We extracted source counts from circular regions with 
40\,arcsec radius from both \nustar modules.  
We identified background regions of 125\,arcsec radius on the same
\nustar detectors as the source.
Since the second and third \nustar observations and the
\chandra observation are nearly contiguous in time and the source is only slowly
varying, we analyzed these data together and refer to them hereafter as the
second epoch.  
We combined the \nustar data from the second and third
observations and from both modules into a single spectrum 
to maximize the signal-to-noise ratio.

We also downloaded and reduced the 13.0\,ksec of \swiftn-XRT PC-mode 
data contemporaneous with the 
first \nustar epoch (\texttt{obsid 00571830006})
using standard procedures in HEASOFT 6.14.  

We processed the \chandra data using standard procedures with CIAO v4.5.
The data were obtained using 1/4 Window readout to reduce pileup;  we
verified that the effect of
pileup on our spectra is negligible and ignore it in further analysis.

We rebinned all of the data to $> 20$ counts per bin and 
fit the data using ISIS v1.6.2-19. 
We also required the
\nustar bins to have SNR of $>4.5$, as above $\sim15$\,keV the background
dominates.
We minimized $\chi^2$ in our fits to the data 
and use
the covariance matrix 
in our significance calculations in Section \ref{sec:fits}.
We used fit energy
bands of 3--30\,keV (\textit{NuSTAR}), 0.3--10\,keV (\textit{Swift}-XRT), 
and 0.2--10\,keV (\textit{Chandra}).
All errors are 90\% C.L., and we have used a cosmology with $h=0.704$,
$\Omega_M=0.273$, $\Omega_\Lambda=0.727$ \citep{Komatsu:11:WMAP7Cosmo}.

\section{Spectral Modeling} \label{sec:fits}

\subsection{Single Power law}
GRB X-ray afterglow spectra are usually well-fit by absorbed power law (PL)
models.
We froze a
Galactic $N_{\rm H}$ component of $1.7\times10^{20}$\,cm$^{-2}$
\citep{Kalberla:05:nH,Evans:2013:GCN15254}
and allowed a varying $N_{\rm H}$ component at the
reported redshift of $z=0.347$. 

A PL fit to the first epoch \nustar data 
shows a clear deficit in the residuals in the 5--6 keV region (Figure
\ref{fig:epoch1fit}).
A joint PL fit including the \swiftn-XRT data improves the parameter
constraints, particularly for $N_{\rm H}$, but the residual structure
remains.
The goodness of fit is poor, with $\chi^2_\nu = 1.6$ (Table \ref{tab:fits}).
A PL fit to the \chandra data and a joint
\nustarn-\chandra PL fit also show residual structure 
(Figure \ref{fig:epoch2fit}) and poor goodness of fit, 
with $\chi^2_\nu=2.2$.  
Additional components (Sections \ref{sec:gabs}--\ref{sec:hardpl}) improve these fits.

\subsection{Absorption Feature} \label{sec:gabs}
\begin{figure}
\includegraphics[width=\columnwidth]{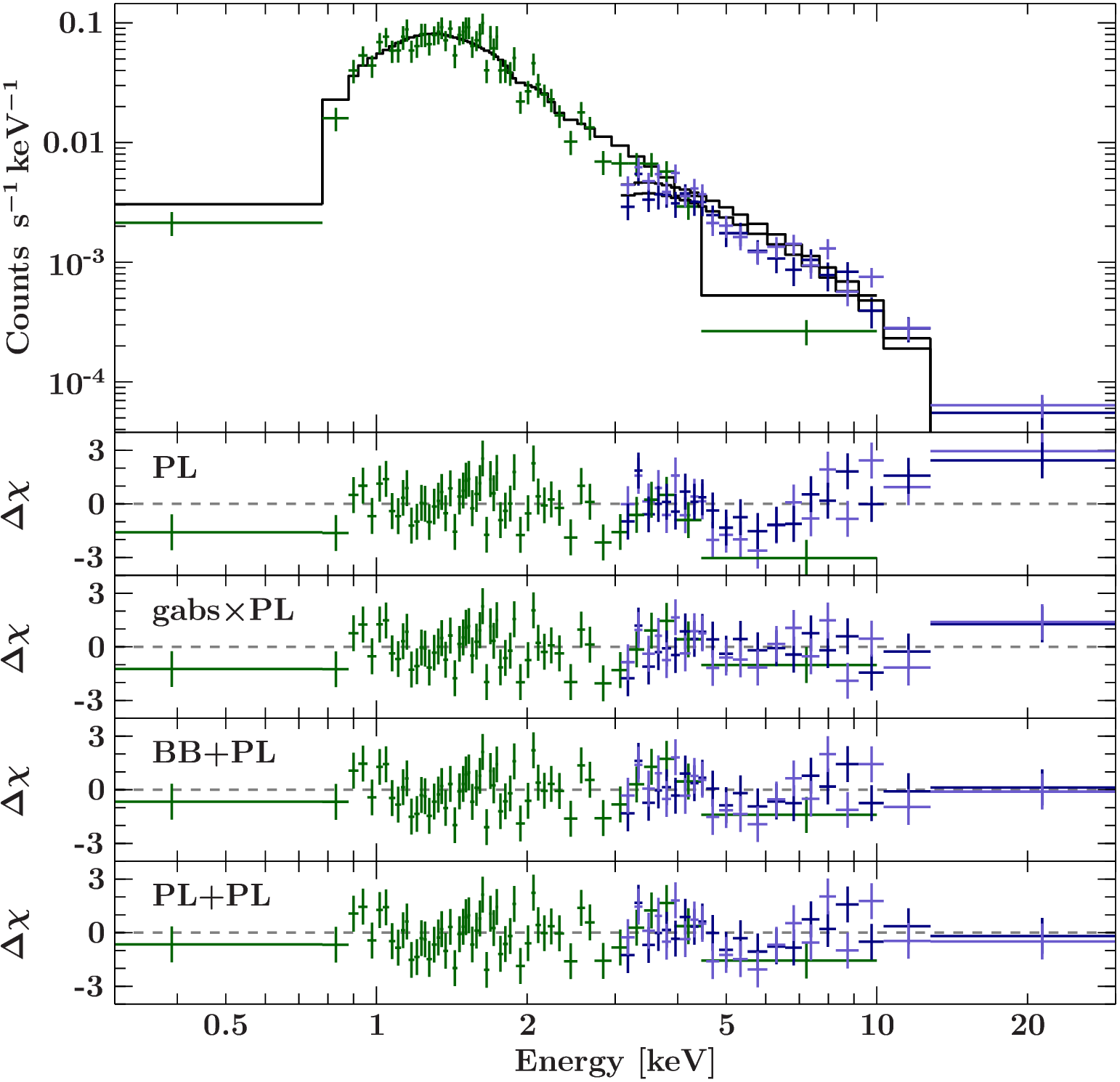}
\caption{Spectral fits to the first epoch \nustar and \swift data. 
The top panel shows the count spectra and PL model fit.
The lower panels show the residuals for the PL, \texttt{gabs}$\times$PL,
BB$+$PL, and PL$+$PL fits. Data are colored blue (\nustar module A), navy blue
(module B)  and green (\swiftn-XRT).
\label{fig:epoch1fit}}
\end{figure}

\begin{figure}
\includegraphics[width=\columnwidth]{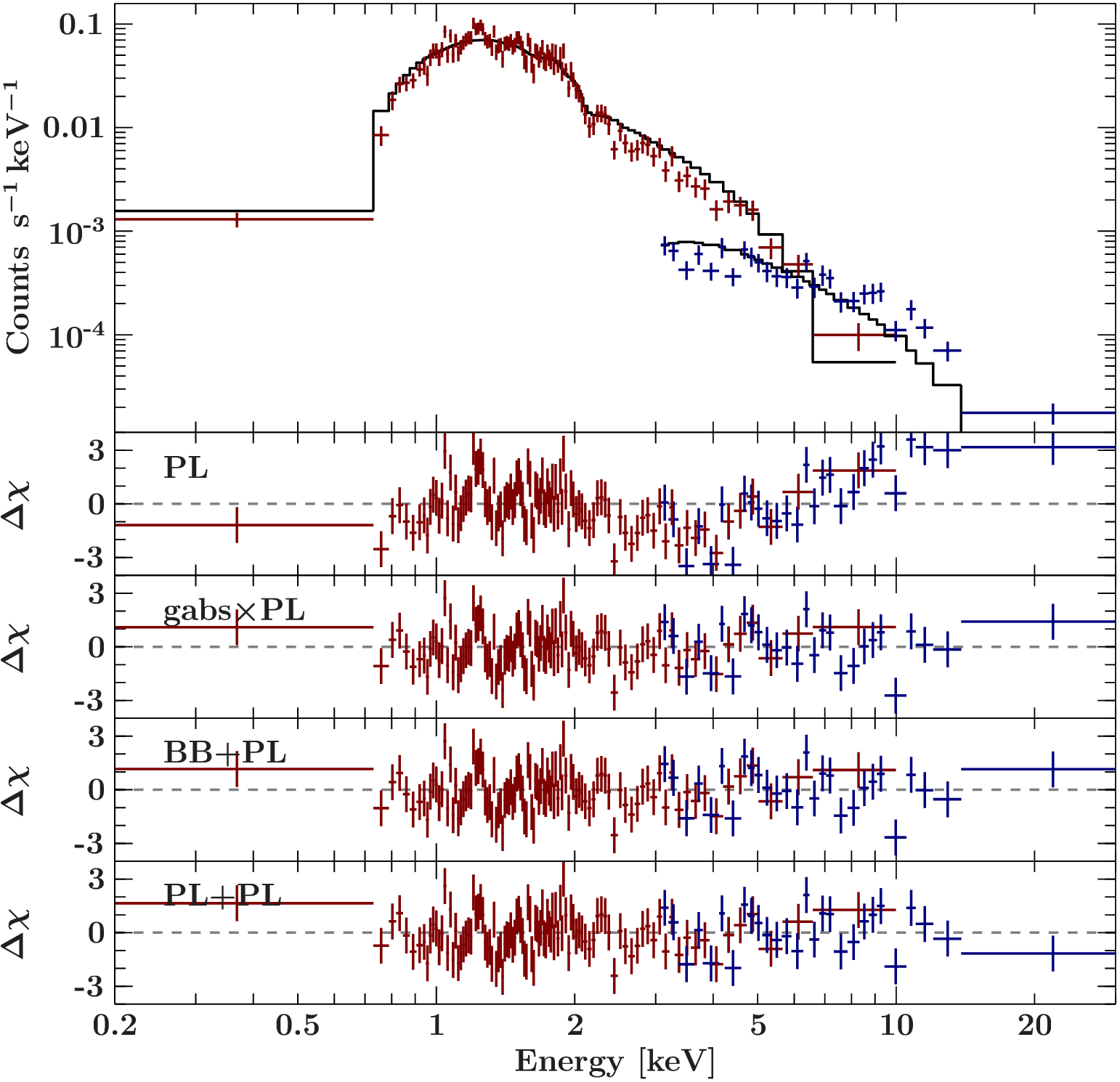}
\caption{Spectral fits to the second epoch \nustar and \chandra data. 
Panels are as in Figure \ref{fig:epoch1fit}.
The \chandra data are 
red, and data from the combined \nustar modules are navy blue.
\label{fig:epoch2fit}}
\end{figure}

Multiplying by a Gaussian absorber (\texttt{gabs}$\times$PL) in the first epoch 
markedly improves the fit residuals relative to a PL fit
(Figure \ref{fig:epoch1fit}). 
The centroid of the Gaussian absorber is at 5.9\spm{0.4}{0.3}\,keV, and
$\sigma=0.9\spm{0.6}{0.3}$\,keV, both in the observer frame.
The \swift data show similar residual structure, and in a joint fit the
Gaussian absorber gives a similar centroid (6.0\spm{0.5}{0.3}\,keV) but
greater width (1.8\spm{1.9}{0.7}\,keV; Table \ref{tab:fits}).
In the joint fit, $\chi^2_\nu$ improves to 1.1 from 1.6 
for three additional parameters.

\begin{figure}
\includegraphics[width=\columnwidth]{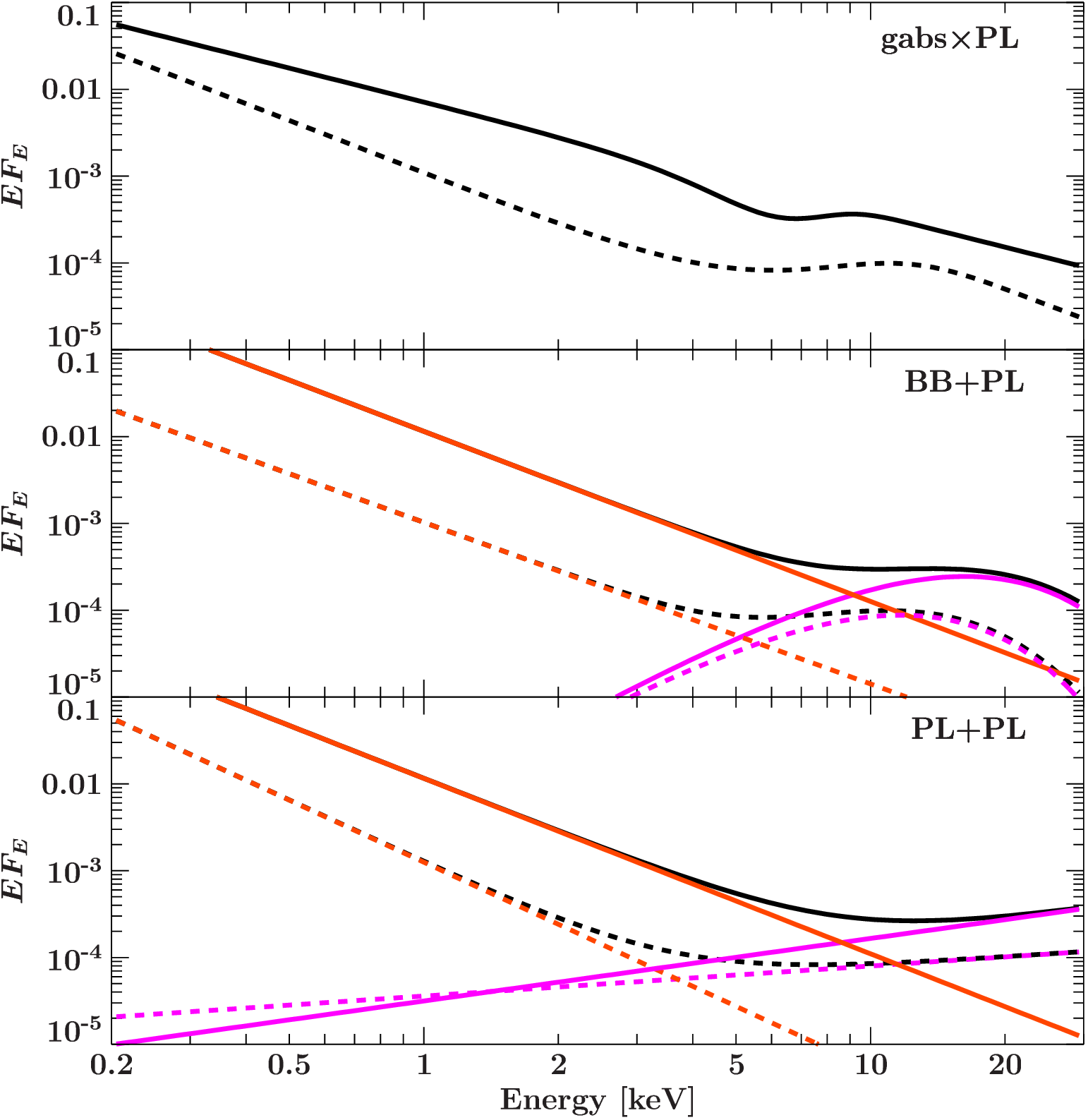}
\caption{Unfolded, unabsorbed model spectra for the
\texttt{gabs}$\times$PL, BB$+$PL, and PL$+$PL fits in
keV$^2$\,cm$^{-2}$\,s$^{-1}$\,keV$^{-1}$.
Fits to the first (second) epoch are plotted with solid (dashed) lines.
For the BB$+$PL and PL$+$PL fits, the total model is plotted in black, the
power law component in orange, and the black body or second power law
component in pink.
\label{fig:nufnu_models}}
\end{figure}

In the second epoch, a Gaussian absorber again improves the fit relative to
a PL ($\chi^2_\nu = 1.2$ from 2.2), but the
parameters are poorly constrained.  The joint \nustar and \chandra fit
provides only an upper limit (4.1\,keV) on the line centroid.
This value is inconsistent with that of
the first epoch, and the required line width is substantially larger 
($\sigma=5.2\spm{2.0}{3.0}$\,keV, Figure \ref{fig:nufnu_models}).  
The large shift in the line centroid
is difficult to explain with absorption by a single species. 
If the large linewidth is interpreted as 
turbulent velocity broadening, this implies relativistic velocities
$\gtrsim0.1c$ that increase from the first epoch to the second, an unlikely
scenario.

\subsection{Bremsstrahlung} We obtained good fits ($\chi^2_\nu \sim 1.1$)
with an absorbed bremsstrahlung plus power-law model (Bremss$+$PL).  The
component is well-constrained in both epochs, with best-fit temperatures of
1.3\ppm{0.2} and 0.83\spm{0.12}{0.11}\,keV in the comoving frame.  The fit
emission measures are 1.1\spm{0.5}{0.3}$\times10^{69}$\,cm$^{-3}$ and
2.3\spm{0.9}{0.6}$\times10^{68}$\,cm$^{-3}$.  These extreme emission
measures, if produced by a constant-density medium, would require densities
of order $10^{10} (R/10^{16}\,\rm{cm})^{-3/2}$ cm$^{-3}$.  However, a
circumstellar medium this dense would be optically thick to electron
scattering, violating the assumptions of the optically thin bremsstrahlung
model.  The emitting region would be optically thin only if the radius of
the region were $> 10^{20}$\,cm, much larger than typical afterglow radii.
More complex density profiles would require even higher densities at some
locations.  Thus while the addition of an optically thin bremsstrahlung
spectral component improves the fit to the data, we are unable to construct
a self-consistent physical interpretation for it.  This problem persists
even if instead we require a higher temperature for the bremsstrahlung
component in order to fit the high-energy excess.  The fit is worse
($\chi^2$ increases by 5.9 in both epochs) and
provides only a lower limit on the temperature ($kT \gtrsim 25$\,keV in the
comoving frame). 
The emission region
must still be larger than $10^{18}$\,cm to be optically thin.

Motivated by the presence of possible additional residual structure in the
\chandra data in the 1--3\,keV range, we attempted to fit 
\texttt{mekal} and \texttt{apec}
plasma emission models
to the second-epoch data.  
With standard
abundances, these models fit metallicity values of zero, reproducing the
unphysical Bremss$+$PL model.  Even with highly variable
abundances, single-temperature plasmas did not provide clear improvements
in the fit.  

\subsection{Black Body}
We also fit a black body plus
power-law model (BB$+$PL).   The $\chi^2$ surface shows two minima for the
blackbody temperature in both epochs, one near 5\,keV and the second near
0.5\,keV.  In the first epoch the higher temperature is preferred
($\chi^2_{\rm low} = 115.3$ vs.\ $\chi^2_{\rm high} = 103.1$ for 90
d.o.f.), while in the second epoch the goodness of fit is closer to
equivalent ($\chi^2_{\rm low} = 156.2$ vs.\ $\chi^2_{\rm high} = 157.7$ for
130 d.o.f).  We argue that the higher-temperature black body fit is more
plausible due to its relative consistency with the component observed in
the first epoch and with theoretical expectations (Section
\ref{sec:discussion}).  

The blackbody components provide 11\% (29\%) 
of the total 0.3--30\,keV flux in the first (second) epoch.  The
implied radii for a spherical emission region
are small and consistent with constant size: 
$1.1\spm{0.5}{0.8}\times10^8$\,cm and $1.5\spm{0.5}{0.6}\times10^{8}$\,cm.
(The radii for the disfavored low-temperature
black bodies are larger but also relatively compact. However, they 
imply a physically unlikely
\emph{contraction} of the emitting region from
$(3.2\ppm{0.8})\times10^{10}$\,cm to $(1.7\ppm{0.4})\times10^{10}$\,cm.)

While black body components have been reported in other GRB afterglow
spectra, none have been observed at such late times, with such high
temperatures, or with such small radii. 
At 1--10\,days
after the burst, the blackbody radius inferred from GRB\,101225A was over
$10^{14}$\,cm and could be explained by the jet interaction with the
circumstellar medium \citep{Thone:11:ChristmasDayBurst}.  
The inferred radius of $10^8$\,cm for \grbn is much harder to explain with
a jet interaction model.
This size
scale is instead on par with the radius of the fallback accretion disks
expected in stellar collapse \citep{Fryer:09:NeutrinoFallback}.  

If we assume we are observing this disk,
the fit temperature can place constraints on the progenitor by
constraining the conditions in the disk.  The luminosity of an accretion
disk is roughly equal to the potential energy released in the accretion.
If we consider material at radius $r$, the luminosity ($L$) is given by
$L=G M_{\rm BH} \dot{m} dr/r^2$ where $\dot{m}$ is the accretion rate and
$dr$ denotes a small annulus of material at radius $r$ (integrating over
$dr$ would produce the total luminosity).  The blackbody emission for such
an annulus is $L = \sigma A T^4 = \sigma 2 \pi r dr$, where $\sigma$ is the
Stefan-Boltzmann constant and $T$ is the blackbody temperature.  If we know
the temperature, we can then derive the accretion rate $\dot{m} = (2 \pi
r^3 \sigma T^4)/(G M_{\rm BH})$.  For our observed temperatures of
4-5.6\,keV, the corresponding accretion rate is $10^{-9}-10^{-10} {\rm
M_\odot s^{-1}}$.  Fallback $10^5-10^6$\,s after a supernova or GRB
explosion has been calculated for a range of progenitors and explosion
energies \citep{MacFadyen:01:Collapsars,tmp_Wong:14:Fallback}.  Fallback at
late times follows a simple power law \citep{Chevalier:89:NSSNAccretion}
and depends on the progenitor and the explosion energy of
the supernova associated with the GRB.  Most fallback calculations
\citep{MacFadyen:01:Collapsars,tmp_Wong:14:Fallback} predict fallback rates
of $10^{-7}-10^{-10} {\rm M_\odot s^{-1}}$ at $10^5-10^6$\,s for supernova
explosions of $1-3\times 10^{51}\,{\rm erg}$.  

Our accretion rates imply a luminosity near $10^5$ times the Eddington
limit for a stellar mass black hole.  Although such extreme super-Eddington
emission rates have been invoked from fallback
\citep{Dexter:13:FallbackLCs}, the exact nature of such transient accretion
is not well known.  Steady-state solutions of disk accretion find that
maintaining emission rates even an order of magnitude above Eddington is
difficult \citep{Jaroszynski:80:BHAccretion}.  Whether such steady state
limits apply in transient situations like our fallback disk remains to be
seen \citep{Abramowicz:05:PolishDonuts}.  Thus without a full model of
these transient events, we are not able to establish a self-consistent
explanation for the blackbody emission.

\subsection{Hard Power Law} \label{sec:hardpl}
Finally, we considered a two power law model (PL$+$PL) 
like that reported for GRB\,111209A \citep{Stratta:13:ULGRB111209A}.
This model is a slightly worse fit in both epochs than the BB$+$PL model
for the same number of free parameters (Table \ref{tab:fits}).

\citet{Stratta:13:ULGRB111209A} interpret the very hard ($\Gamma \sim 0$)
second PL component they report for GRB\,111209A at 70\,ksec after the
burst as the tail of the hard
power-law emission sometimes observed by \textit{Fermi}-LAT \citep[e.g.,][and
references therein]{Zhang:11:LATGRBs}.  
This component is detected in the late
prompt and early afterglow phases and decays according to a powerlaw; its
physical origin remains uncertain.  The
non-detection by LAT of both GRBs complicates this interpretation.  An
extrapolation of our Epoch\,1 powerlaw flux to the 0.1--10\,GeV band gives a
photon flux of $3\times10^{-6}$\,photons\,cm$^{-2}$\,s$^{-1}$, a value
higher than the upper limit of
$7\times10^{-7}$\,photons\,cm$^{-2}$\,s$^{-1}$ reported by
\citet{Kocevski:2013:GCN15268} in the first 2\,ksec after the burst, when
the afterglow---and thus presumably the hard component---was much brighter.
The problem is even more severe for the component reported by
\citet{Stratta:13:ULGRB111209A}: its higher flux and much harder spectral
index extrapolate to a 0.1--10\,GeV photon flux of
1.5\,photons\,cm$^{-2}$\,s$^{-1}$, an extremely high value sufficient to
trigger LAT.  We examined the late-time LAT data for both bursts and
confirm no excess emission.  Consistency with the nondetection by LAT in
both cases thus requires a cutoff above the \nustar and \textit{XMM}
bandpasses but below the the LAT bandpass at 30\,MeV.  This
phenomenological model is plausible, but the connection of these components
to the early-time hard power-law components detected by LAT in other GRBs 
therefore remains speculative.

\begin{deluxetable*}{|l|cccc|cccc|}
\tablecolumns{7} 
\tablewidth{0pc} 
\tablecaption{Best-fit parameters of spectral models. Errors are 90\% C.L. \label{tab:fits}}
\tablehead{ 
\colhead{}    
& \multicolumn{4}{c}{\nustar \& \swift Epoch 1} 
& \multicolumn{4}{c}{\nustar \& \chandra Epoch 2}  \\
\colhead{Parameter} 
& \colhead{PL} & \colhead{\texttt{gabs}$\times$PL} & \colhead{BB$+$PL} &
\colhead{PL$+$PL}
& \colhead{PL} & \colhead{\texttt{gabs}$\times$PL} & \colhead{BB$+$PL} &
\colhead{PL$+$PL}
}
\startdata
$N_{\rm H}$ ($10^{22}$ cm$^{-2}$)  
	& 2.55\spm{0.24}{0.23} & 2.71\spm{0.41}{0.30} & 3.32\spm{0.35}{0.34} &
	3.35\spm{0.40}{0.37}
	& 1.98\ppm{0.14} & 2.82\spm{0.31}{0.439} & 2.74\spm{0.21}{0.20} &
	3.02\spm{0.29}{0.27} \\

$\Gamma$
	& 3.33\ppm{0.13} & 3.29\spm{0.17}{0.14} & 3.96\spm{0.24}{0.23} & 4.02\spm{0.33}{3.47}
	& 3.06\ppm{0.11} & 3.95\spm{0.55}{0.87} & 3.86\spm{0.20}{0.19} &
	4.37\spm{0.40}{0.12} \\

$\Gamma_2$
	& & & & 1.28\spm{0.65}{0.73}
	& & & & 1.65\spm{0.26}{0.29}  \\

$E_0$ (keV)
	& & 5.89\spm{0.53}{0.33}  & &
	& & $< 4.09$ &  & \\

$\sigma$ (keV) 
	& & 1.75\spm{1.97}{0.70}  & &
	& & 5.15\spm{1.97}{2.97} & & \\

$\tau(E = E_0)$
	& & 0.73\spm{0.12}{0.18} & & 
	& & 2.8\spm{1.3}{1.8}  & & \\

$kT$ (comoving frame, keV) 
	& &  & 5.58\spm{2.12}{1.15} &
	& &  & 4.02\spm{0.73}{0.56} & \\
$\chi^2$/$\nu$ 
	& 146.8/92  & 98.1/89 & 103.1/90 & 105.9/90
 	& 288.8/132  & 158.9/129  & 157.7/130 & 161.0/130 \\
$P_\chi(X>\chi | \nu) $ 
	&  2.5E-4 &  0.23 & 0.16  & 0.12
	& 1.0E-13 &  0.04 &  0.05 & 0.03

\enddata
\end{deluxetable*}

\subsection{Component significance}
We verified the significance of the additional spectral components using
Monte Carlo simulations according to the method of posterior predictive
p-values \citep[ppp-values;][]{Protassov:02:LRTest}.  We initialized each
fit by stepping the additional feature through a grid in energy and finding
the largest relative improvement in $\chi^2$
\citep[c.f.][]{Hurkett:08:XRTLineSearches}.  This procedure accounts for
the ``look-elsewhere'' effect of multiple trials, as we have no \textit{a
priori} expectation of the observed line energy or component temperature.
In none of our $10^4$ simulated realizations of a null PL model did fits
with alternative models (\texttt{gabs}$\times$PL, BB$+$PL, or PL$+$PL) produce
improvements in $\chi^2$ as large as observed in the real data.  This
implies that the spectral features are significant at $>3.9\sigma$ in both
epochs: the $\chi^2$ improvement for each model fit is extremely unlikely
to be due to chance if the true underlying model were simply an absorbed
PL.  

\section{Conclusion} \label{sec:discussion}

Our late-time afterglow observations of \grb require an additional spectral
component at high significance.  Several alternative models provide 
acceptable fits to the data.
These spectral features are detected more than 1\,Msec after the burst,
much later than any components previously reported in X-ray afterglows, 
probing a largely unexplored phase of afterglow evolution.
Several unique features of \grb make it possible to detect these late-time
features for the first time.  The unusually bright 
afterglow enables high-quality spectral fits, and \nustar has excellent
sensitivity at the relevant energies and can constrain the continuum above
10 keV.  Moreover, the primary power law is unusually soft, so
the high-energy component is not swamped.  It is not
yet clear whether this emission is related to progenitor physics
unique to this unusual, ultra-long burst; \nustar observations of the 
bright ``canonical'' long GRB\,130427A were consistent with emission by a
single spectral component \citep{Kouveliotou:13:GRB130427A}.  Future
observations of bright afterglows will be needed to determine the
prevalence of these late-time spectral components and identify the relevant
emission mechanism.

\acknowledgments
This work was supported under NASA Contract No. NNG08FD60C and uses
data from the \nustar mission, a project led by the California Institute of
Technology, managed by the Jet Propulsion Laboratory, and funded by the
National Aeronautics and Space Administration. We thank the \nustar
Operations team for executing the ToO observations.
This research has used the 
\nustar Data Analysis Software (NuSTARDAS) jointly developed by the ASI
Science Data Center (ASDC, Italy) and the California Institute of
Technology (USA).
These results are based in part on observations made by the \chandra X-ray
Observatory.
We thank the \chandra director for granting discretionary time and the
\chandra team for prompt execution of the observations.

{\it Facilities:} \facility{NuSTAR}, \facility{Swift}, \facility{Chandra}.

\clearpage

\begin{thebibliography}{48}
\expandafter\ifx\csname natexlab\endcsname\relax\def\natexlab#1{#1}\fi

\bibitem[{{Abramowicz}(2005)}]{Abramowicz:05:PolishDonuts}
{Abramowicz}, M.~A. 2005, \href{http://dx.doi.org/10.1007/11403913_49}{in
  Growing Black Holes: Accretion in a Cosmological Context, ed. A.~{Merloni},
  S.~{Nayakshin}, \& R.~A. {Sunyaev}}, 257

\bibitem[{{Amati} {et~al.}(2000){Amati}, {Frontera}, {Vietri}, {in't Zand},
  {Soffitta}, {Costa}, {Del Sordo}, {Pian}, {Piro}, {Antonelli}, {Fiume},
  {Feroci}, {Gandolfi}, {Guidorzi}, {Heise}, {Kuulkers}, {Masetti},
  {Montanari}, {Nicastro}, {Orlandini}, \& {Palazzi}}]{Amati:00:AGAbsLine}
{Amati}, L., {Frontera}, F., {Vietri}, M., {et~al.} 2000,
  \href{http://dx.doi.org/10.1126/science.290.5493.953}{Science, 290, 953}

\bibitem[{{Bellm} {et~al.}(2013){Bellm}, {Harrison}, {Forster}, {Madsen},
  {Rana}, {Boggs}, {Tomsick}, {Miller}, \& {Cenko}}]{Bellm:13:GCN15286}
{Bellm}, E.~C., {Harrison}, F.~A., {Forster}, K., {et~al.} 2013, GRB
  Coordinates Network, 15286, 1

\bibitem[{{Bloom} {et~al.}(2011){Bloom}, {Giannios}, {Metzger}, {Cenko},
  {Perley}, {Butler}, {Tanvir}, {Levan}, {O'Brien}, {Strubbe}, {De Colle},
  {Ramirez-Ruiz}, {Lee}, {Nayakshin}, {Quataert}, {King}, {Cucchiara},
  {Guillochon}, {Bower}, {Fruchter}, {Morgan}, \& {van der
  Horst}}]{Bloom:11:SwiftJ1644TDF}
{Bloom}, J.~S., {Giannios}, D., {Metzger}, B.~D., {et~al.} 2011,
  \href{http://dx.doi.org/10.1126/science.1207150}{Science, 333, 203}

\bibitem[{{Burrows} {et~al.}(2013){Burrows}, {Malesani}, {Lien}, {Cenko}, \&
  {Gehrels}}]{Burrows:2013:GCN15253}
{Burrows}, D.~N., {Malesani}, D., {Lien}, A.~Y., {Cenko}, S.~B., \& {Gehrels},
  N. 2013, GRB Coordinates Network, 15253, 1

\bibitem[{{Burrows} {et~al.}(2011){Burrows}, {Kennea}, {Ghisellini}, {Mangano},
  {Zhang}, {Page}, {Eracleous}, {Romano}, {Sakamoto}, {Falcone}, {Osborne},
  {Campana}, {Beardmore}, {Breeveld}, {Chester}, {Corbet}, {Covino},
  {Cummings}, {D'Avanzo}, {D'Elia}, {Esposito}, {Evans}, {Fugazza}, {Gelbord},
  {Hiroi}, {Holland}, {Huang}, {Im}, {Israel}, {Jeon}, {Jeon}, {Jun}, {Kawai},
  {Kim}, {Krimm}, {Marshall}, {P.~M{\'e}sz{\'a}ros}, {Negoro}, {Omodei},
  {Park}, {Perkins}, {Sugizaki}, {Sung}, {Tagliaferri}, {Troja}, {Ueda},
  {Urata}, {Usui}, {Antonelli}, {Barthelmy}, {Cusumano}, {Giommi}, {Melandri},
  {Perri}, {Racusin}, {Sbarufatti}, {Siegel}, \&
  {Gehrels}}]{Burrows:11:SwiftJ1644TDF}
{Burrows}, D.~N., {Kennea}, J.~A., {Ghisellini}, G., {et~al.} 2011,
  \href{http://dx.doi.org/10.1038/nature10374}{\nat, 476, 421}

\bibitem[{{Campana} {et~al.}(2006){Campana}, {Mangano}, {Blustin}, {Brown},
  {Burrows}, {Chincarini}, {Cummings}, {Cusumano}, {Della Valle}, {Malesani},
  {M{\'e}sz{\'a}ros}, {Nousek}, {Page}, {Sakamoto}, {Waxman}, {Zhang}, {Dai},
  {Gehrels}, {Immler}, {Marshall}, {Mason}, {Moretti}, {O'Brien}, {Osborne},
  {Page}, {Romano}, {Roming}, {Tagliaferri}, {Cominsky}, {Giommi}, {Godet},
  {Kennea}, {Krimm}, {Angelini}, {Barthelmy}, {Boyd}, {Palmer}, {Wells}, \&
  {White}}]{Campana:06:ThermalAG}
{Campana}, S., {Mangano}, V., {Blustin}, A.~J., {et~al.} 2006,
  \href{http://dx.doi.org/10.1038/nature04892}{\nat, 442, 1008}

\bibitem[{{Cenko} {et~al.}(2012){Cenko}, {Krimm}, {Horesh}, {Rau}, {Frail},
  {Kennea}, {Levan}, {Holland}, {Butler}, {Quimby}, {Bloom}, {Filippenko},
  {Gal-Yam}, {Greiner}, {Kulkarni}, {Ofek}, {Olivares E.}, {Schady},
  {Silverman}, {Tanvir}, \& {Xu}}]{Cenko:12:TDESwiftJ2058}
{Cenko}, S.~B., {Krimm}, H.~A., {Horesh}, A., {et~al.} 2012,
  \href{http://dx.doi.org/10.1088/0004-637X/753/1/77}{\apj, 753, 77}

\bibitem[{{Chevalier}(1989)}]{Chevalier:89:NSSNAccretion}
{Chevalier}, R.~A. 1989, \href{http://dx.doi.org/10.1086/168066}{\apj, 346,
  847}

\bibitem[{{Dexter} \& {Kasen}(2013)}]{Dexter:13:FallbackLCs}
{Dexter}, J., \& {Kasen}, D. 2013,
  \href{http://dx.doi.org/10.1088/0004-637X/772/1/30}{\apj, 772, 30}

\bibitem[{{Evans} {et~al.}(2013){Evans}, {Pagani}, {Page}, {Beardmore},
  {Starling}, \& {Lien}}]{Evans:2013:GCN15254}
{Evans}, P.~A., {Pagani}, C., {Page}, K.~L., {et~al.} 2013, GRB Coordinates
  Network, 15254, 1

\bibitem[{{Fitzpatrick} \& {the Fermi GBM
  Team}(2013)}]{Fitzpatrick:2013:GCN15255}
{Fitzpatrick}, G., \& {the Fermi GBM Team}. 2013, GRB Coordinates Network,
  15255, 1

\bibitem[{{Friis} \& {Watson}(2013)}]{Friis:13:AGBBSearch}
{Friis}, M., \& {Watson}, D. 2013,
  \href{http://dx.doi.org/10.1088/0004-637X/771/1/15}{\apj, 771, 15}

\bibitem[{{Fryer}(2009)}]{Fryer:09:NeutrinoFallback}
{Fryer}, C.~L. 2009, \href{http://dx.doi.org/10.1088/0004-637X/699/1/409}{\apj,
  699, 409}

\bibitem[{{Gendre} {et~al.}(2013){Gendre}, {Stratta}, {Atteia}, {Basa},
  {Bo{\"e}r}, {Coward}, {Cutini}, {D'Elia}, {Howell}, {Klotz}, \&
  {Piro}}]{Gendre:13:ULGRB111209A}
{Gendre}, B., {Stratta}, G., {Atteia}, J.~L., {et~al.} 2013,
  \href{http://dx.doi.org/10.1088/0004-637X/766/1/30}{\apj, 766, 30}

\bibitem[{{Golenetskii} {et~al.}(2013){Golenetskii}, {Aptekar}, {Frederiks},
  {Pal\'shin}, {Oleynik}, {Ulanov}, \& {et al.}}]{Golenetskii:2013:GCN15260}
{Golenetskii}, S., {Aptekar}, R., {Frederiks}, D., {et~al.} 2013, GRB
  Coordinates Network, 15260, 1

\bibitem[{{Harrison} {et~al.}(2013){Harrison}, {Craig}, {Christensen},
  {Hailey}, {Zhang}, {Boggs}, {Stern}, {Cook}, {Forster}, {Giommi},
  {Grefenstette}, {Kim}, {Kitaguchi}, {Koglin}, {Madsen}, {Mao}, {Miyasaka},
  {Mori}, {Perri}, {Pivovaroff}, {Puccetti}, {Rana}, {Westergaard}, {Willis},
  {Zoglauer}, {An}, {Bachetti}, {Barri{\`e}re}, {Bellm}, {Bhalerao},
  {Brejnholt}, {Fuerst}, {Liebe}, {Markwardt}, {Nynka}, {Vogel}, {Walton},
  {Wik}, {Alexander}, {Cominsky}, {Hornschemeier}, {Hornstrup}, {Kaspi},
  {Madejski}, {Matt}, {Molendi}, {Smith}, {Tomsick}, {Ajello}, {Ballantyne},
  {Balokovi{\'c}}, {Barret}, {Bauer}, {Blandford}, {Brandt}, {Brenneman},
  {Chiang}, {Chakrabarty}, {Chenevez}, {Comastri}, {Dufour}, {Elvis}, {Fabian},
  {Farrah}, {Fryer}, {Gotthelf}, {Grindlay}, {Helfand}, {Krivonos}, {Meier},
  {Miller}, {Natalucci}, {Ogle}, {Ofek}, {Ptak}, {Reynolds}, {Rigby},
  {Tagliaferri}, {Thorsett}, {Treister}, \& {Urry}}]{Harrison:2013:NuSTAR}
{Harrison}, F.~A., {Craig}, W.~W., {Christensen}, F.~E., {et~al.} 2013,
  \href{http://dx.doi.org/10.1088/0004-637X/770/2/103}{\apj, 770, 103}

\bibitem[{{Hurkett} {et~al.}(2008){Hurkett}, {Vaughan}, {Osborne}, {O'Brien},
  {Page}, {Beardmore}, {Godet}, {Burrows}, {Capalbi}, {Evans}, {Gehrels},
  {Goad}, {Hill}, {Kennea}, {Mineo}, {Perri}, \&
  {Starling}}]{Hurkett:08:XRTLineSearches}
{Hurkett}, C.~P., {Vaughan}, S., {Osborne}, J.~P., {et~al.} 2008,
  \href{http://dx.doi.org/10.1086/586881}{\apj, 679, 587}

\bibitem[{{Jaroszynski} {et~al.}(1980){Jaroszynski}, {Abramowicz}, \&
  {Paczynski}}]{Jaroszynski:80:BHAccretion}
{Jaroszynski}, M., {Abramowicz}, M.~A., \& {Paczynski}, B. 1980, Acta
Astron., 30, 1

\bibitem[{{Kalberla} {et~al.}(2005){Kalberla}, {Burton}, {Hartmann}, {Arnal},
  {Bajaja}, {Morras}, \& {P{\"o}ppel}}]{Kalberla:05:nH}
{Kalberla}, P.~M.~W., {Burton}, W.~B., {Hartmann}, D., {et~al.} 2005,
  \href{http://dx.doi.org/10.1051/0004-6361:20041864}{\aap, 440, 775}

\bibitem[{{Kocevski} {et~al.}(2013){Kocevski}, {Racusin}, {Vianello},
  {Axelsson}, \& {Omodei}}]{Kocevski:2013:GCN15268}
{Kocevski}, D., {Racusin}, J., {Vianello}, G., {Axelsson}, M., \& {Omodei}, N.
  2013, GRB Coordinates Network, 15268, 1

\bibitem[{{Komatsu} {et~al.}(2011){Komatsu}, {Smith}, {Dunkley}, {Bennett},
  {Gold}, {Hinshaw}, {Jarosik}, {Larson}, {Nolta}, {Page}, {Spergel},
  {Halpern}, {Hill}, {Kogut}, {Limon}, {Meyer}, {Odegard}, {Tucker}, {Weiland},
  {Wollack}, \& {Wright}}]{Komatsu:11:WMAP7Cosmo}
{Komatsu}, E., {Smith}, K.~M., {Dunkley}, J., {et~al.} 2011,
  \href{http://dx.doi.org/10.1088/0067-0049/192/2/18}{\apjs, 192, 18}

\bibitem[{{Kouveliotou} {et~al.}(2013){Kouveliotou}, {Granot}, {Racusin},
  {Bellm}, {Vianello}, {Oates}, {Fryer}, {Boggs}, {Christensen}, {Craig},
  {Dermer}, {Gehrels}, {Hailey}, {Harrison}, {Melandri}, {McEnery}, {Mundell},
  {Stern}, {Tagliaferri}, \& {Zhang}}]{Kouveliotou:13:GRB130427A}
{Kouveliotou}, C., {Granot}, J., {Racusin}, J.~L., {et~al.} 2013,
  \href{http://dx.doi.org/10.1088/2041-8205/779/1/L1}{\apjl, 779, L1}

\bibitem[{{Levan} {et~al.}(2011){Levan}, {Tanvir}, {Cenko}, {Perley},
  {Wiersema}, {Bloom}, {Fruchter}, {Postigo}, {O'Brien}, {Butler}, {van der
  Horst}, {Leloudas}, {Morgan}, {Misra}, {Bower}, {Farihi}, {Tunnicliffe},
  {Modjaz}, {Silverman}, {Hjorth}, {Th{\"o}ne}, {Cucchiara}, {Cer{\'o}n},
  {Castro-Tirado}, {Arnold}, {Bremer}, {Brodie}, {Carroll}, {Cooper}, {Curran},
  {Cutri}, {Ehle}, {Forbes}, {Fynbo}, {Gorosabel}, {Graham}, {Hoffman},
  {Guziy}, {Jakobsson}, {Kamble}, {Kerr}, {Kasliwal}, {Kouveliotou},
  {Kocevski}, {Law}, {Nugent}, {Ofek}, {Poznanski}, {Quimby}, {Rol},
  {Romanowsky}, {S{\'a}nchez-Ram{\'{\i}}rez}, {Schulze}, {Singh}, {van
  Spaandonk}, {Starling}, {Strom}, {Tello}, {Vaduvescu}, {Wheatley}, {Wijers},
  {Winters}, \& {Xu}}]{Levan:11:SwiftJ1644TDF}
{Levan}, A.~J., {Tanvir}, N.~R., {Cenko}, S.~B., {et~al.} 2011,
  \href{http://dx.doi.org/10.1126/science.1207143}{Science, 333, 199}

\bibitem[{{Levan} {et~al.}(2013){Levan}, {Tanvir}, {Starling}, {Wiersema},
  {Page}, {Perley}, {Schulze}, {Wynn}, {Chornock}, {Hjorth}, {Cenko},
  {Fruchter}, {O'Brien}, {Brown}, {Tunnicliffe}, {Malesani}, {Jakobsson},
  {Watson}, {Berger}, {Bersier}, {Cobb}, {Covino}, {Cucchiara}, {de Ugarte
  Postigo}, {Fox}, {Gal-Yam}, {Goldoni}, {Gorosabel}, {Kaper}, {Kruehler},
  {Karjalainen}, {Osborne}, {Pian}, {Sanchez-Ramirez}, {Schmidt}, {Skillen},
  {Tagliaferri}, {Thone}, {Vaduvescu}, {Wijers}, \&
  {Zauderer}}]{tmp_Levan:13:ULGRBs}
{Levan}, A.~J., {Tanvir}, N.~R., {Starling}, R.~L.~C., {et~al.} 2013, ArXiv
  e-prints, \href{http://arxiv.org/abs/1302.2352}{{\sffamily arXiv:1302.2352
  [astro-ph.HE]}}

\bibitem[{{Lien} {et~al.}(2013){Lien}, {Markwardt}, {Page}, {Palmer},
  {Racusin}, {Siegel}, \& {Ukwatta}}]{Lien:2013:GCN15246}
{Lien}, A.~Y., {Markwardt}, C.~B., {Page}, K.~L., {et~al.} 2013, GRB
  Coordinates Network, 15246, 1

\bibitem[{{MacFadyen} {et~al.}(2001){MacFadyen}, {Woosley}, \&
  {Heger}}]{MacFadyen:01:Collapsars}
{MacFadyen}, A.~I., {Woosley}, S.~E., \& {Heger}, A. 2001,
  \href{http://dx.doi.org/10.1086/319698}{\apj, 550, 410}

\bibitem[{{Markwardt} {et~al.}(2013){Markwardt}, {Barthelmy}, {Baumgartner},
  {Cummings}, {Fenimore}, {Gehrels}, \& {et al.}}]{Markwardt:2013:GCN15257}
{Markwardt}, C.~B., {Barthelmy}, S.~D., {Baumgartner}, W.~H., {et~al.} 2013,
  GRB Coordinates Network, 15257, 1

\bibitem[{{Nakauchi} {et~al.}(2013){Nakauchi}, {Kashiyama}, {Suwa}, \&
  {Nakamura}}]{Nakauchi:13:BSGULGRB}
{Nakauchi}, D., {Kashiyama}, K., {Suwa}, Y., \& {Nakamura}, T. 2013,
  \href{http://dx.doi.org/10.1088/0004-637X/778/1/67}{\apj, 778, 67}

\bibitem[{{Piro} {et~al.}(2000){Piro}, {Garmire}, {Garcia}, {Stratta}, {Costa},
  {Feroci}, {M{\'e}sz{\'a}ros}, {Vietri}, {Bradt}, {Frail}, {Frontera},
  {Halpern}, {Heise}, {Hurley}, {Kawai}, {Kippen}, {Marshall}, {Murakami},
  {Sokolov}, {Takeshima}, \& {Yoshida}}]{Piro:2000:AGLine}
{Piro}, L., {Garmire}, G., {Garcia}, M., {et~al.} 2000,
  \href{http://dx.doi.org/10.1126/science.290.5493.955}{Science, 290, 955}

\bibitem[{{Protassov} {et~al.}(2002){Protassov}, {van Dyk}, {Connors},
  {Kashyap}, \& {Siemiginowska}}]{Protassov:02:LRTest}
{Protassov}, R., {van Dyk}, D.~A., {Connors}, A., {Kashyap}, V.~L., \&
  {Siemiginowska}, A. 2002, \href{http://dx.doi.org/10.1086/339856}{\apj, 571,
  545}

\bibitem[{{Reeves} {et~al.}(2002){Reeves}, {Watson}, {Osborne}, {Pounds},
  {O'Brien}, {Short}, {Turner}, {Watson}, {Mason}, {Ehle}, \&
  {Schartel}}]{Reeves:02:AGLine}
{Reeves}, J.~N., {Watson}, D., {Osborne}, J.~P., {et~al.} 2002,
  \href{http://dx.doi.org/10.1038/416512a}{\nat, 416, 512}

\bibitem[{{Sako} {et~al.}(2005){Sako}, {Harrison}, \&
  {Rutledge}}]{Sako:05:AGLineReanalysis}
{Sako}, M., {Harrison}, F.~A., \& {Rutledge}, R.~E. 2005,
  \href{http://dx.doi.org/10.1086/425644}{\apj, 623, 973}

\bibitem[{{Savchenko} {et~al.}(2013){Savchenko}, {Beckmann}, {Ferrigno},
  {Bozzo}, {Beck}, {Borkowski}, \& {et al.}}]{Savchenko:2013:GCN15259}
{Savchenko}, V., {Beckmann}, V., {Ferrigno}, C., {et~al.} 2013, GRB Coordinates
  Network, 15259, 1

\bibitem[{{Sparre} \& {Starling}(2012)}]{Sparre:12:AGBBSearch}
{Sparre}, M., \& {Starling}, R.~L.~C. 2012,
  \href{http://dx.doi.org/10.1111/j.1365-2966.2012.21858.x}{\mnras, 427, 2965}

\bibitem[{{Starling} {et~al.}(2012){Starling}, {Page}, {Pe'Er}, {Beardmore}, \&
  {Osborne}}]{Starling:12:AGBBody}
{Starling}, R.~L.~C., {Page}, K.~L., {Pe'Er}, A., {Beardmore}, A.~P., \&
  {Osborne}, J.~P. 2012,
  \href{http://dx.doi.org/10.1111/j.1365-2966.2012.22116.x}{\mnras, 427, 2950}

\bibitem[{{Stratta} {et~al.}(2013){Stratta}, {Gendre}, {Atteia}, {Bo{\"e}r},
  {Coward}, {De Pasquale}, {Howell}, {Klotz}, {Oates}, \&
  {Piro}}]{Stratta:13:ULGRB111209A}
{Stratta}, G., {Gendre}, B., {Atteia}, J.~L., {et~al.} 2013,
  \href{http://dx.doi.org/10.1088/0004-637X/779/1/66}{\apj, 779, 66}

\bibitem[{{Sudilovsky} {et~al.}(2013{\natexlab{a}}){Sudilovsky}, {Kann},
  {Greiner}, \& {GROND team}}]{Sudilovsky:2013:GCN15247}
{Sudilovsky}, V., {Kann}, D., {Greiner}, J., \& {GROND team}.
  2013{\natexlab{a}}, GRB Coordinates Network, 15247, 1

\bibitem[{{Sudilovsky} {et~al.}(2013{\natexlab{b}}){Sudilovsky}, {Kann},
  {Schady}, {Klose}, {Greiner}, \& {Kruehler}}]{Sudilovsky:2013:GCN15250}
{Sudilovsky}, V., {Kann}, D., {Schady}, P., {et~al.} 2013{\natexlab{b}}, GRB
  Coordinates Network, 15250, 1

\bibitem[{{Suzuki} \& {Shigeyama}(2013)}]{Suzuki:13:ThermalCocoon}
{Suzuki}, A., \& {Shigeyama}, T. 2013,
  \href{http://dx.doi.org/10.1088/2041-8205/764/1/L12}{\apjl, 764, L12}

\bibitem[{{Suzuki} {et~al.}(2013){Suzuki}, {Sakakibara}, {Negoro}, {Nakahira},
  {Tomida}, {Ueno}, \& {et al.}}]{Suzuki:2013:GCN15248}
{Suzuki}, K., {Sakakibara}, H., {Negoro}, H., {et~al.} 2013, GRB Coordinates
  Network, 15248, 1

\bibitem[{{Tanvir} {et~al.}(2013){Tanvir}, {Levan}, {Hounsell}, {Fruchter},
  {Cenko}, {Perley}, \& {O'Brien}}]{Tanvir:13:GCN15489}
{Tanvir}, N.~R., {Levan}, A.~J., {Hounsell}, R., {et~al.} 2013, GRB Coordinates
  Network, 15489, 1

\bibitem[{{Th{\"o}ne} {et~al.}(2011){Th{\"o}ne}, {de Ugarte Postigo}, {Fryer},
  {Page}, {Gorosabel}, {Aloy}, {Perley}, {Kouveliotou}, {Janka}, {Mimica},
  {Racusin}, {Krimm}, {Cummings}, {Oates}, {Holland}, {Siegel}, {de Pasquale},
  {Sonbas}, {Im}, {Park}, {Kann}, {Guziy}, {Garc{\'{\i}}a}, {Llorente},
  {Bundy}, {Choi}, {Jeong}, {Korhonen}, {Kub{\`a}nek}, {Lim}, {Moskvitin},
  {Mu{\~n}oz-Darias}, {Pak}, \& {Parrish}}]{Thone:11:ChristmasDayBurst}
{Th{\"o}ne}, C.~C., {de Ugarte Postigo}, A., {Fryer}, C.~L., {et~al.} 2011,
  \href{http://dx.doi.org/10.1038/nature10611}{\nat, 480, 72}

\bibitem[{{Vreeswijk} {et~al.}(2013){Vreeswijk}, {Malesani}, {Fynbo}, {De Cia},
  \& {Ledoux}}]{Vreeswijk:2013:GCN15249}
{Vreeswijk}, P.~M., {Malesani}, d., {Fynbo}, J.~P.~U., {De Cia}, A., \&
  {Ledoux}, C. 2013, GRB Coordinates Network, 15249, 1

\bibitem[{{Wong} {et~al.}(2014){Wong}, {Fryer}, {Ellinger}, {Rockefeller}, \&
  {Kalogera}}]{tmp_Wong:14:Fallback}
{Wong}, T.-W., {Fryer}, C.~L., {Ellinger}, C.~I., {Rockefeller}, G., \&
  {Kalogera}, V. 2014, ArXiv e-prints,
  \href{http://arxiv.org/abs/1401.3032}{{\sffamily arXiv:1401.3032
  [astro-ph.HE]}}

\bibitem[{{Woosley} \& {Heger}(2012)}]{Woosley:12:LongCollapsars}
{Woosley}, S.~E., \& {Heger}, A. 2012,
  \href{http://dx.doi.org/10.1088/0004-637X/752/1/32}{\apj, 752, 32}

\bibitem[{{Zhang} {et~al.}(2013){Zhang}, {Zhang}, {Murase}, {Connaughton}, \&
  {Briggs}}]{tmp_Zhang:13:ULGRBs}
{Zhang}, B.-B., {Zhang}, B., {Murase}, K., {Connaughton}, V., \& {Briggs},
  M.~S. 2013, ArXiv e-prints, \href{http://arxiv.org/abs/1310.2540}{{\sffamily
  arXiv:1310.2540 [astro-ph.HE]}}

\bibitem[{{Zhang} {et~al.}(2011){Zhang}, {Zhang}, {Liang}, {Fan}, {Wu},
  {Pe'er}, {Maxham}, {Gao}, \& {Dong}}]{Zhang:11:LATGRBs}
{Zhang}, B.-B., {Zhang}, B., {Liang}, E.-W., {et~al.} 2011,
  \href{http://dx.doi.org/10.1088/0004-637X/730/2/141}{\apj, 730, 141}

\end{thebibliography}
\end{document}